\documentclass[11pt,english,epsf]{scrartcl}
\usepackage{lmodern}
\usepackage[T1]{fontenc}
\usepackage[latin9]{inputenc}
\usepackage{geometry}
\usepackage{amsmath}
\usepackage{amssymb}

\newtheorem{theorem}{Theorem}

\newcommand {\dfn} {\stackrel{\Delta} {=}}

\newcommand {\ta} {\tilde{a}}

\newcommand {\bx} {\mbox{\boldmath $x$}}

\newcommand{\calS}{{\cal S}}

\newcommand{\calU}{{\cal U}}

\newcommand{\calX}{{\cal X}}
\newcommand{\calY}{{\cal Y}}

\allowdisplaybreaks

\begin{document}
\thispagestyle{empty}
\title{On Empirical Cumulant Generating Functions of Code Lengths for
Individual Sequences}
%\thanks{This research was supported by my wife and kids.}
\author{Neri Merhav
%\thanks{
%Currently on sabbatical leave at HP Laboratories,
%1501 Page Mill Road, MS 3U-4, Palo Alto CA 94304, USA.}
}
\date{}
\maketitle

\begin{center}
Department of Electrical Engineering \\
Technion - Israel Institute of Technology \\
Technion City, Haifa 32000, ISRAEL \\
E--mail: {\tt merhav@ee.technion.ac.il}\\
\end{center}
\vspace{1.5\baselineskip}
\setlength{\baselineskip}{1.5\baselineskip}

\begin{center}
{\bf Abstract}
\end{center}
\setlength{\baselineskip}{0.5\baselineskip}
We consider the problem of lossless compression of individual sequences using
finite--state (FS) machines, from the perspective of the best achievable empirical
cumulant generating function (CGF) of the code length, i.e., the normalized logarithm of the empirical
average of the exponentiated code length. Since the probabilistic
CGF is minimized in terms of the R\'enyi entropy of the source,
one of the motivations of this study is
to derive an individual--sequence analogue of the R\'enyi entropy,
in the same way that the FS compressibility is the
individual--sequence counterpart of the Shannon entropy.
We consider the CGF of the code-length both from the perspective of
fixed--to--variable (F-V) length coding and the perspective of
variable--to--variable (V-V) length coding, where the latter turns out to yield a
better result, that coincides with the FS compressibility.
We also extend our results to compression with side information, available at
both the encoder and decoder. In this case, the V--V version no longer
coincides with the FS compressibility, but results in a different complexity
measure.\\

\vspace{0.2cm}
{\bf Index Terms} Individual sequences, compressibility, finite--state
machines, cumulant generating function, R\'enyi entropy, Lempel--Ziv
algorithm.

\setlength{\baselineskip}{2\baselineskip}
\newpage

\section{Introduction}

The celebrated paper by Ziv and Lempel \cite{ZL78} was one of the earliest
works (if not the first) in the information theory literature that adopted the
individual--sequence approach as an alternative to the traditional
probabilistic approach (see, e.g., \cite[Sections III, IV]{MF98} and many
references therein).
In the context of lossless source coding, according to this approach, the system model
imposes certain limitations
on the resources of the encoder (which is modeled as a finite--state machine) rather than
on the statistics of the source sequence to be compressed. One of the most
important concepts contributed in \cite{ZL78} was the notion of {\it
finite--state compressibility} of a given infinite source sequence, i.e., the best compression ratio achievable
by any finite--state (FS) machine that may compress this sequence. The importance of
the FS compressibility is rooted in the fact that it is the
individual--sequence analogue of the notion of the entropy rate: while the
entropy rate is an asymptotically achievable lower 
bound on the minimum normalized expected code length 
in the probabilistic scenario, the FS compressibility is an
asymptotically tight lower bound on the minimum normalized empirical
expectation of the code length (achievable by FS encoders) in the
individual--sequence setting. Moreover, the FS compressiblity of a realization
of a finite--alphabet, stationary and ergodic process is equal to the entropy rate
almost surely \cite[Theorem 4]{ZL78}.

Turning for a moment to the traditional probabilistic setting, 
it is well known that,
while the normalized expected code length (or equivalently, 
the expected compression ratio) has always been the most customary
figure of merit, other figure of merits
for compression have also been raised in the literature. Most importantly,
the cumulant generating function (CGF) of the code length, namely, the
normalized logarithm of the exponential moment of the code-length\footnote{A
more precise definition will follow in the sequel.} was first proposed by
Campbell \cite{Campbell65}
as a performance criterion for lossless compression, along with a corresponding coding
theorem in terms of the R\'enyi entropy.
Campbell's motivation was that the CGF enhances more strongly the contribution of the
longest codewords (even if they are weighted by small probabilities), and so, the
resulting code optimization is more conservative since the code length
fluctuations tend to be reduced. In the realm of stochastic control
(and referring to more general problems with any utility function, not
necessarily just code length),
such a property
is called {\it risk--sensitivity}, and accordingly, the CGF cost function
is called a {\it risk--sensitive} cost. Additional motivations for the CGF
of a cost function in general, include: (i) robustness against uncertainty in the source statistics, (ii)
optimization of the {\it full}
distribution of the cost (in some cases),
and not just the first moment, and (iii) intimate relationship to the large deviations
performance (via the Chernoff bound), which in the context of data
compression, has
implications on design considerations concerning the buffer overflow probability, see, e.g.,
\cite{Humblet81}, \cite{Jelinek68}, \cite{Merhav91}, \cite{UH99},
\cite{Wyner74}. It is also intimately related to the problem of guessing
\cite{Arikan96}.
For a somewhat more elaborate discussion on
risk--sensitive cost functions, see, e.g., \cite[Introduction]{CISpaper} and many
references therein.

Combining the contents of the above two paragraphs together, it is now natural to
raise the question of what can be said about the individual--sequence counterpart
of the CGF of the code--length, namely, 
the empirical CGF of 
the code length. In other words, we wish to find an achievable
lower bound on the normalized logarithm of the empirical average of an
exponential function of the code--length. Such an achievable lower bound would then play a
role in a natural
definition of an individual--sequence analogue of the R\'enyi
entropy, in parallel to the analogy between the FS compressibility and 
the Shannon entropy rate. It should be noted, however, that there is an
important difference between the Shannon entropy and the R\'enyi entropy, in
this context.
While the Shannon entropy rate, defined by the limit of normalized joint
entropies, always exists for
a stationary ergodic source, there is no established R\'enyi entropy rate for
such a process in general, as the corresponding limit does not always exist.
For this reason, there will be no attempt to take this limit, i.e., our
results will be stated in terms of a given finite order (or block length).

The above verbal description of the empirical CGF was deliberately given
somewhat vaguely, because there is some freedom in the choice of the exact definition, and
different definitions turn out to yield different results. Indeed, in the sequel, we
will consider a few definitions, and characterize the corresponding achievable
lower bounds. In most cases, the achievability will be accomplished by some
variant of the Lempel--Ziv (LZ) algorithm \cite{ZL78}.
In particular, we will consider the CGF of the code-length both from the
perspective of 
fixed--to--variable (F-V) length coding and the perspective of
variable--to--variable (V-V) length coding,
where the latter turns out to yield a
better result, that coincides with the FS compressibility.
We also extend our results to compression with side information, available at
both the encoder and decoder. In this case, the V--V version no longer
coincides with the FS compressibility, but results in a different complexity
measure.

The remaining part of this paper is organized as follows. 
In Section 2, is the setup is first formulated, and then it is divided
into two subsections, the first being 
devoted to the class of F--V empirical CGFs
and the second one -- to V--V empirical CGFs. In Section 3, our main
results are extended to a situation of coding with side information.
Finally, in Section 4, the main findings of this work are summarized.

\section{Problem Formulation and Main Results}

We begin by reviewing the model of a finite--state encoder of
Ziv and Lempel \cite{ZL78}. Let $\bx=(x_1,x_2,\ldots)$ be a deterministic, infinite source
sequence (individual sequence) to be compressed, where each $x_i$ takes values
in a finite alphabet $\calX$ of size $\alpha$.
An $s$--state
encoder $E$ is defined by a quintuple
$(\calS,\calX,\calY,f,g)$, where
$\calS$ is a finite set
of $s$ states, $\calX$ is the finite source alphabet just described,
$\calY$ is
a finite set of binary words
(possibly of different lengths, including the null word for idling),
$f:\calS\times\calX\to\calY$ is the encoder output function, and
$g:\calS\times\calX\to\calS$ is the next--state function.
When the input sequence $(x_1,x_2,...)$ is fed
sequentially into the encoder $E$, the
latter outputs a sequence of binary words $(y_1,y_2,\ldots)$, $y_i\in\calY$,
while going through a sequence of states $(z_1,z_2,\ldots)$,
$z_i\in\calS$,
according to
\begin{equation}
\label{fsm}
y_i=f(z_i,x_i),~~~z_{i+1}=g(z_i,x_i),~~~i=1,2,...
\end{equation}
where $z_i$ is the state of encoder $E$ at time instant $i$.
The decoder, on the other hand, receives the sequence $y_1,y_2,\ldots$
and reconstructs the source sequence $x_1,x_2,\ldots$.
In the sequel, we will use the conventional shorthand notation $x_i^j$ for
the string segment $(x_i,x_{i+1},\ldots,x_j)$ whenever $j \ge i$. For $i=1$,
we will omit the subscript $i$ and denote $(x_1,x_2,\ldots,x_j)$ by $x^j$.
Similar rules will apply to other sequences, like the state sequence and the
encoder output sequence. 
As in \cite{ZL78}, in the sequel, we will use the shorthand
notation $f(z_1,x^n)$ and $g(z_1,x^n)$ for the output $y^n=(y_1,\ldots,y_n)$
and the sequence of states $z^n=(z_1,\ldots,z_n)$ that are obtained as the
response of $E$
to $x^n=(x_1,\ldots,x_n)$ for a given initial state $z_1$.

Following the terminology of \cite{ZL78}, a FS encoder $E$
is said to be {\it information
lossless} (IL) if for all $z_1\in\calS$ and all
$x^n\in\calX^n$, the triple $(z_1,f(z_1,x^n),g(z_1,x^n))$ uniquely determines $x^n$.
The length function
associated with $E$ is defined as 
\begin{equation}
L_E(y^n)=\sum_{i=1}^nl(y_i),
\end{equation}
where $l(y_i)$ is the length of the binary string $y_i\in\calY$,
which may include the option of $l(y_i)=0$ for the 
null output, which is the case when the
encoder is idling, i.e., waiting for additional inputs before producing further
compressed output bits.

While the compression ratio $L_E(y^n)/n$ can be viewed as the empirical
expectation of the code lengths 
$\{l(y_i)\}$, in this work, we are focusing on the empirical
expectations of several exponential functions of these lengths, viewing them
as individual--sequence counterparts of the ordinary
probabilistic expectations of these
functions. It turns out that there is considerable freedom 
in the definition of this kind of figure
of merit, and the corresponding optimal codes are sensitive to the exact
definition. 

\subsection{Fixed--to--Variable Length CGFs and the Empirical R\'enyi
Entropy}

For a
given $\lambda> 0$, the simplest objective of this kind is the quantity
\begin{equation}
\label{1st}
\frac{1}{\lambda}\log\left[\frac{1}{n}\sum_{t=1}^n
2^{\lambda l(y_t)}\right],
\end{equation}
where here and throughout the sequel, logarithms are defined to the base 2.
The problem with this objective function is that many data compression
algorithms (with block codes 
as well as Lempel--Ziv algorithms included) work in ``bursts''.
In other words, most of the time they idle (which means $l(y_t)=0$) and
only in relatively few time instants they actually 
output chunks of compressed bits.
The undesirable property of the objective (\ref{1st})
is that each time instant of such an idling stage contributes
a term of $2^{\lambda\cdot 0}$ to the
sum $\sum_{t=1}^n 2^{\lambda l(y_t)}$, and so, there is overall an additive
term, which is almost as large as $n\cdot 2^{\lambda\cdot 0}=n$, even though
the code length contributed at these times is zero. 
One possible remedy to this undesired property is to simply ignore these terms.
Another possibility is to define
the empirical average of an exponential function of the code length for 
an $\ell$--block, and so, when $\ell$ is large enough, it is conceivable that
at least one $l(y_t)$ within each block is positive, and even 
if this is not the
case, the
seemingly superfluous term of $2^{\lambda\cdot 0}=1$ 
is added only once in a block,
rather than almost each time instant (and so, the relative contribution would be
insignificant).
Consider then the more general objective function
\begin{equation}
\label{f2v}
\frac{1}{\lambda\ell}\log_2\left[\frac{\ell}{n}\sum_{t=0}^{n/\ell-1}
2^{\lambda L(y_{t\ell+1}^{t\ell+\ell})}\right],
\end{equation}
where it is assumed that $\ell$ is a positive integer that divides $n$.
We next present a simple result 
concerning the objective (\ref{f2v}).

\begin{theorem}
For every IL encoder with $s$ states,
\begin{equation}
\label{renyilb}
\frac{1}{\lambda\ell}\log_2\left[\frac{\ell}{n}\sum_{t=0}^{n/\ell-1}
2^{\lambda L(y_{t\ell+1}^{t\ell+\ell})}\right]\ge
\hat{H}_\lambda^\ell(x^n)-\frac{\gamma(s,\ell)}{\ell},
\end{equation}
where 
\begin{equation}
\hat{H}_\lambda^\ell(x^n)=\frac{1+\lambda}{\lambda\ell}\log_2\left(
\sum_{a^\ell\in\calX^\ell}
[\hat{P}(a^\ell)]^{1/(1+\lambda)}\right),
\end{equation}
$\hat{P}(a^\ell)$ being the empirical probability (relative frequency) of
$a^\ell\in\calX^\ell$ in $x^n$ along its $n/\ell$ non--overlapping
$\ell$--blocks,
$\{x_{t\ell+1}^{t\ell+\ell}\}$, and
\begin{equation}
\gamma(s,\ell)=2\log
s+\log\left[1+\log\left(\frac{s^2+\alpha^\ell}{s^2}\right)\right].
\end{equation}
\end{theorem}
The first term on the r.h.s.\ of eq.\ (\ref{renyilb}) is the empirical $\ell$--th order
R\'enyi entropy associated with $x^n$, which is the natural
individual--sequence counterpart of the ordinary $\ell$--th order R\'enyi
entropy of the probabilistic setting. The second term expresses (an estimate
of) the extra compression capability allowed by the memory of the FS encoder
(captured in its state $z_i$), which may carry useful information between the
successive blocks. When $\ell \gg \log s$, however, this extra compression capability
becomes relatively negligible,
because the amount of past information memorized by the state is very small
compared to the amount of information in each source block of size $\ell$.
The lower bound of Theorem 1 can be essentially achieved by applying a Shannon code
for $\ell$ blocks, which is matched to the probability distribution that is proportional to
$[\hat{P}(a^\ell)]^{1/(1+\lambda)}$, and appending a header 
of size about $|\calX|^\ell\log(n/\ell+1)$, 
describing the empirical distribution
$\{\hat{P}(a^\ell),~a^\ell\in\calX^\ell\}$ (i.e., the type information). 
Since this is a logarithmic function of $n$, this overhead redundancy vanishes
for large $n$.
However, there is
still a gap here in the sense that the number of states required to implement such
an encoder is by far larger than all values of $s$ that keep
$\gamma(s,\ell)/\ell$ reasonably small for a given $\ell$.

{\it Proof.}
For a given IL encoder $E$, let
$L'(a^\ell)=\min_{z_1\in\calS}L[f(z_1,a^\ell)]$, and define the probability
distribution
\begin{equation}
Q(a^\ell)=\frac{2^{-L'(a^\ell)}}{\sum_{\ta^\ell}2^{-L'(\ta^\ell)}}
\end{equation}
Then,
\begin{equation}
L'(a^\ell)=-\log Q(a^\ell)-\log \left[\sum_{\ta^\ell}2^{-L'(\ta^\ell)}\right]\ge
-\log Q(a^\ell)-\gamma(s,\ell)
\end{equation}
where the second inequality is supported by Lemma 2 of \cite{ZL78} (the
generalized Kraft inequality). Thus,
\begin{eqnarray}
\frac{\ell}{n}\sum_{t=0}^{n/\ell-1}
\exp_2\{\lambda L(y_{t\ell+1}^{t\ell+\ell})\}&=&
\frac{\ell}{n}\sum_{t=0}^{n/\ell-1}
\exp_2\{\lambda L[f(z_{t\ell+1},x_{t\ell+1}^{t\ell+\ell})]\}\nonumber\\
&\ge& \frac{\ell}{n}\sum_{t=0}^{n/\ell-1}
\exp_2\{\lambda L'(x_{t\ell+1}^{t\ell+\ell})\}\nonumber\\
&\ge&\frac{\ell}{n}\sum_{t=0}^{n/\ell-1}
\exp_2\{\lambda[-\log Q(x_{t\ell+1}^{t\ell+\ell})-\gamma(s,\ell)]\}\nonumber\\
&=&2^{-\lambda\gamma(s,\ell)}\sum_{a^\ell}\frac{\hat{P}(a^\ell)}{Q^\lambda(a^\ell)}\nonumber\\
&\ge&2^{-\lambda\gamma(s,\ell)}\left(\sum_{a^\ell}
[\hat{P}(a^\ell)]^{1/(1+\lambda)}\right)^{1+\lambda}\nonumber\\
&\ge&\exp_2\{\lambda[\ell\hat{H}_\lambda^\ell(x^n)-\gamma(s,\ell)]\},
\end{eqnarray}
where the second to the last inequality is obtained by minimizing the
expression $\sum_{a^\ell}\hat{P}(a^\ell)/Q^\lambda(a^\ell)$ w.r.t.\ the
probability distribution $Q$. Finally, the desired result
is obtained by taking the base 2 logarithm of both sides and
the normalizing by $\lambda\ell$. This completes the proof of Theorem 1.
$\Box$

For very large $\ell$, there is another (conceptually) simple lower bound that is essentially
attained by applying the LZ78 algorithm to each
$\ell$--block separately, namely, restarting the LZ dictionary at every time
instant $t$ which is an integer multiple of $\ell$. By Theorem 1 of
\cite{ZL78}, we know that for any $s$--state IL encoder,
$L(y_{t\ell+1}^{t\ell+\ell})$ is lower bounded by
$(c_t+s^2)\log\frac{c_t+s^2}{4s^2}$, where
$c_t$ is the maximum number of distinct phrases in
$x_{t\ell+1}^{t\ell+\ell}$, and so,
\begin{eqnarray}
& &\frac{1}{\lambda\ell}\log\left[\frac{\ell}{n}\sum_{t=0}^{n\/\ell-1}\exp_2\{\lambda
L(y_{t\ell+1}^{t\ell+\ell})\}\right]\nonumber\\
&\ge&\frac{1}{\lambda\ell}\log\left[\frac{\ell}{n}\sum_{t=0}^{n/\ell-1} \exp_2\left\{\lambda
(c_t+s^2)\log[(c_t+s^2)/4s^2]\right\}\right].
\end{eqnarray}
The second line can be thought of as an alternative definition
of the R\'enyi
counterpart of the compressibility of individual sequences.

\subsection{Variable--to--Variable Length CGFs and the LZ Complexity}

The problem with the objective function 
(\ref{f2v}) is that for large $\ell$ and large $\lambda$, the performance
becomes extremely sensitive to fluctuations in the code lengths,
$L(y_{t\ell+1}^{(t+1)\ell})$.
Clearly, for a given average of
$\{L(y_{t\ell+1}^{(t+1)\ell})\}$, 
eq.\ (\ref{f2v}) is minimized when
all lengths are equal to this average (as can easily be understood from
Jensen's inequality), namely, when the fluctuations are completely eliminated.
This observation motivates us to expand the scope and redefine our objective
in the spirit of 
variable--to--variable length coding which allows much more freedom
in the quest for reducing the length fluctuations.

Specifically, rather than the above segmentation of the source string 
$x^n$ into fixed--length
blocks of size $\ell$,
consider a sequence--dependent segmentation 
according to a set (or dictionary) of $c$ distinct variable--length strings, which all have
(at least approximately) the same empirical probability, in other words,
the empirical distribution of this set of strings is uniform, or nearly
uniform. In such a case, it would make sense that, at least in the 
absence of constraints on the encoder structure, the code lengths for
those strings would be all the same (or nearly so), and then the
length fluctuations would be eliminated altogether. 
For the class of FS encoders
considered here, we may not be able to guarantee uniform lengths always, but this can
certainly serve at least as a guideline for good code design.

A natural way to accomplish such a segmentation with a
uniform empirical distribution, is by parsing the sequence into
$c$ {\it distinct} phrases, in the spirit of the parsings described in 
\cite{ZL78}. In this case, every such phrase (or
string) appears exactly once, and so, its empirical probability is $1/c$.
Accordingly, for a given $x^n$ and a given
parsing the sequence into $c$ distinct\footnote{With the possible
exception of the last phrase, which may be incomplete.} phrases,
$x_1^{n_1},x_{n_1+1}^{n_2},\ldots,x_{n_{c-1}+1}^n$, we define
\begin{equation}
\rho_E^\lambda(x^n)=\frac{c}{n\lambda}\log\left[\frac{1}{c}\sum_{i=1}^c
2^{\lambda L(y_{n_{i-1}+1}^{n_i})}\right],~
~~~n_0\equiv 0,~~n_c\equiv n,
\end{equation}
where the factor $c/n$ outside the logarithm is meant to normalize the
empirical CGF by the average phrase length, $n/c$, in analogy to the factor
of $1/\ell$ outside the logarithm in eq.\ (\ref{f2v}).

Informally speaking, had the dictionary of the 
various phrases been known in advance to both encoder and
decoder, then ideally (i.e., ignoring the finite--state structure of the encoder),
the compressed form of each one of these phrases would be
of length $L(y_{n_{i-1}+1}^{n_i})=\log c$, and
hence, intuitively, one would expect that essentially, $\rho_E^\lambda(x^n)$ 
cannot be smaller than
\begin{equation}
\label{clogc}
\frac{c}{n\lambda}\log\left[\frac{1}{c}\sum_{i=1}^c2^{\lambda
\log c}\right]=\frac{c\log c}{n},
\end{equation}
which is also the main term of the lower bound on the ordinary compressibility
(see \cite{ZL78}). In other words, it seems plausible that
$\sum_{i=1}^c
\exp_2\{\lambda L(y_{n_{i-1}+1}^{n_i})\}$ (which lacks the normalization by
$c$) should be lower bounded by an expression
whose exponential order is as large as $2^{(\lambda+1)\log c}$.
The next theorem supports this intuition more formally.

\begin{theorem}
Given an arbitrary IL encoder $E$ with no more than $s$
states, and given a source sequence $x^n$ with $c$ distinct phrases,
\begin{equation}
\sum_{i=1}^c
\exp_2\{\lambda L(y_{n_{i-1}+1}^{n_i})\}\ge
\frac{s^2\left[\exp_2\left\{(\lambda+1)\log\left(\frac{c+s^2}{2s^2}\right)
\right\}-1\right]}
{2^{\lambda+1}-1}.
\end{equation}
\end{theorem}

\vspace{0.1cm}

\noindent
{\it Proof.}
Given $x^n$ and its parsing into $c$ different phrases,
let $c_j$ denote the number of phrases for which the total compressed bit string is of
length $j$, that is, $L(y_{n_{i-1}+1}^{n_i})=
\sum_{t=n_{i-1}+1}^{n_i}l(y_t)=j$. As argued in \cite{ZL78}, the IL property
of the encoder implies that $c_j\le
s^22^j$ for all $j$, because the initial state, the final state, and the compressed
sequence in between uniquely determine the source string.
As is also argued in \cite{ZL78}, in order to derive a lower bound, 
one may assume ideal packing of
minimal lengths and thereby overestimate
$c_j$ as $s^22^j$ for $j=0,1,\ldots,k$, where $k$ is the largest integer such
that $c \ge \sum_{j=0}^k s^22^j=s^2(2^{k+1}-1)$, which means that
$c < s^2(2^{k+2}-1)$.
Thus,
\begin{eqnarray}
\sum_{i=1}^c
2^{\lambda L(y_{n_{i-1}+1}^{n_i})}&\ge&
s^2\sum_{j=0}^k 2^{\lambda j}\cdot 2^j\nonumber\\
&=&\frac{s^2[2^{(k+1)(\lambda+1)}-1]}{2^{\lambda+1}-1}.
\end{eqnarray}
But from the above definition of $k$, we have 
\begin{equation}
k\ge
\log\left(\frac{c+s^2}{s^2}\right)-2=\log\left(\frac{c+s^2}{2s^2}\right)-1,
\end{equation}
and so,
\begin{equation}
\sum_{i=1}^c
2^{\lambda L(y_{n_{i-1}+1}^{n_i})}\ge
\frac{s^2\left[\exp_2\left\{(\lambda+1)\log\left(\frac{c+s^2}{2s^2}\right)
\right\}-1\right]}
{2^{\lambda+1}-1},
\end{equation}
which completes the proof of Theorem 2. $\Box$

An alternative lower bound can be obtained using the same technique as in the
lower bound in Subsection 2.1,
where instead of averaging w.r.t.\ the empirical distribution of
non--overlapping $\ell$--blocks,
$\{\hat{P}(x^\ell)\}$, as was done in Subsection 2.1, here we have
the uniform empirical distribution $\hat{P}(w)=1/c$,
where $\{w\}$ are the $c$ distinct phrases.
In this case, Lemma 2 of \cite{ZL78}
(the generalized
Kraft inequality) applies too, but with
$\alpha^\ell$ being replaced by $c$ in the definition of $\gamma(s,\ell)$,
i.e., here the logarithm of the
Kraft sum, $\log\left[\sum_{w} 2^{-L'(w)}\right]$,
is upper bounded by $\gamma(s,\log_\alpha c)$.
The resulting alternative to the lower bound of Theorem 2 would then be
\begin{equation}
\label{alternativelb}
\sum_{i=1}^c
\exp_2\{\lambda L(y_{n_{i-1}+1}^{n_i})\}\ge \exp_2\{(\lambda+1)\log c
-\lambda\gamma(s,\log_\alpha c)\}.
\end{equation}
Here too, the leading term at the exponent of the lower bound is
$(\lambda+1)\log c$. None of the two lower bounds dominates the other, in
general. The answer to the question which one is tighter depends on the parameters of
the problem.

A compatible upper bound is now established for the case where the $c$ phrases
are obtained by the incremental parsing procedure of \cite{ZL78}, according to
which $x^n$ is phrased sequentially, where each new phrase is the shortest
string not encountered before as a phrase.

\begin{theorem}
Let $x^n$ be given and let $c$ denote the number of
phrases resulting from the incremental parsing procedure.
Let $L_{\mbox{\tiny LZ}}(x_{n_{i-1}+1}^{n_i})$ denote the total length associated with the
compression of the $i$--th phrase according to the LZ78 algorithm \cite{ZL78}. Then,
\begin{equation}
\sum_{i=1}^c\exp_2\{\lambda L_{\mbox{\tiny LZ}}(x_{n_{i-1}+1}^{n_i})\}\le
(2\alpha)^{\lambda}2^{(\lambda+1)\log c}.
\end{equation}
\end{theorem}

The theorem tells that the LZ78 algorithm essentially achieves the lower bound
of Theorem 2 (for this choice of $c$) 
in the sense that the exponential order 
of the upper bound (as an exponential function of $\log c$)
is the same as that of the lower bound, as they both behave like $2^{(\lambda+1)\log
c}$ in their leading term, and uniformly for every $\lambda > 0$. Accordingly,
the above--mentioned lower bound on
$\rho_E^\lambda(x^n)$, which is about $\frac{c\log c}{n}$, is asymptotically
achieved in Theorem 3. It is interesting to observe that although the LZ78 algorithm behaves like a
variable--to--variable length code (as it maps variable--length source phrases
into variable--length compressed bit strings), it achieves essentially the same
performance as that of the ideal variable--to--fixed length code described before,
which is, as said, free of the undesirable length fluctuations. Moreover, unlike that ideal
variable--to--fixed length code, which is aware of the dictionary of
phrases in advance, the LZ78 algorithm achieves this performance without
knowing this dictionary ahead of time, and independendtly of $\lambda$.

\noindent
{\it Proof.}
We refer the reader to the proof of Theorem 2 in \cite{ZL78}.
Let $x^{n_1},x_{n_1+1}^{n_2},x_{n_2+1}^{n_3},\ldots,x_{n_c+1}^n$ denote the
phrases that result from the incremental parsing procedure. As described in
the constructive proof of \cite[Theorem 2]{ZL78} (which describes 
the LZ78 algorithm), the $i$--th phrase $x_{n_{i-1}+1}^{n_i}$
is encoded by $L_{\mbox{\tiny LZ}}(x_{n_{i-1}+1}^{n_i})=\lceil\log(\alpha i)\rceil$ bits. Thus,
\begin{eqnarray}
\sum_{i=1}^c
\exp_2\{\lambda L_{\mbox{\tiny LZ}}(x_{n_{i-1}+1}^{n_i})\}&=&
\sum_{i=1}^c \exp_2\{\lambda \lceil \log(\alpha i)\rceil\}\nonumber\\
&\le&\sum_{i=1}^c \exp_2\{\lambda[\log(\alpha i)+1]\}\nonumber\\
&\le&c\cdot \exp_2\{\lambda[\log(\alpha c)+1]\}
=(2\alpha)^{\lambda}\cdot 2^{(\lambda+1)\log c},
\end{eqnarray}
completing the proof of Theorem 3. $\Box$

\section{Extension to Coding with Side Information}

We now extend our main results to the case where side information is available
to both the encoder and decoder. We begin by re-formulating the FS encoder model 
so as to allow access to side information. 

An $s$-state
encoder $E$ with side information is defined by a set of six objects,
$(\calS,\calX,\calY,\calU,f,g)$, where
$\calS$, $\calX$ and $\calY$ are as before,
$\calU$ is a finite alphabet
of side information,
$f:\calS\times\calX\times\calU\to\calY$ is the encoder output function, and
$g:\calS\times\calX\times\calU\to\calS$ is the next--state function.
When an input sequence $(x_1,x_2,\ldots)$ and a side information
sequence $(u_1,u_2,\ldots)$ are fed
together, sequentially into $E$, the
encoder outputs a sequence of binary words $(y_1,y_2,\ldots)$, 
while going through a sequence of states $(z_1,z_2,...)$,
$z_i\in\calS$,
according to
\begin{equation}
\label{fsm}
y_i=f(z_i,x_i,u_i),~~~z_{i+1}=g(z_i,x_i,u_i),~~~i=1,2,...
\end{equation}
where $z_i$ is the state of $E$ at time instant $i$.
The decoder receives the pair sequence
$(y_1,u_1),(y_2,u_2),\ldots$
and reconstructs the source sequence $(x_1,x_2,\ldots)$.

A finite--state encoder $E$
with side information is said to be {\it information
lossless} (IL) if for every $z_1\in\calS$ and all
$(x^n,u^n)\in\calX^n\times\calU^n$, $n\ge 1$,
the quadruple $(z_1,z_{n+1},y^n,u^n)$ uniquely determines $x^n$,
where $z_{n+1}$ and $y^n=(y_1,\ldots,y_n)$ are obtained by iterating
eq.\ (\ref{fsm}) with $z_1$,
$x^n$, and $u^n$ as inputs. As before, the length function
associated with $E$ is defined as $L_E(y^n)=\sum_{i=1}^nl(y_i)$.

As for the fixed--to--variable CGF, one can easily extend the
derivation in Subsection 2.1 as follows.
\begin{eqnarray}
\frac{\ell}{n}\sum_{t=0}^{n/\ell-1}\exp_2\{\lambda
L(y_{t\ell+1}^{t\ell+\ell})\}&\ge&
\sum_{u^\ell}\hat{P}(u^\ell)\cdot
2^{\lambda[\hat{H}_\lambda^\ell(x^n|u^\ell)-\gamma(s,\ell)]}\nonumber\\
&\dfn&2^{-\lambda\gamma(s,\ell)}\sum_{u^\ell}\hat{P}(u^\ell)\left\{\sum_{x^\ell}
[\hat{P}(x^\ell|u^\ell)]^{1/(1+\lambda)}\right\}^{1+\lambda}\nonumber\\
&=&2^{-\lambda\gamma(s,\ell)}\sum_{u^\ell}\left\{\sum_{x^\ell}
[\hat{P}(x^\ell,u^\ell)]^{1/(1+\lambda)}\right\}^{1+\lambda},
\end{eqnarray}
whose main factor is related to the empirical conditional R\'enyi 
entropy of order $\ell$.

For the variable--to--variable CGF, following \cite{Ziv85},
consider a certain parsing of the sequence of pairs
$(x_1,u_1),(x_2,u_2),\ldots,(x_n,u_n)$,
into $c\equiv c(x^n,u^n)$ distinct phrases.
Let $c(u^n)$ be the number of distinct phrases of $u^n$
and let $c_k(x^n|u^n)$ be the number of distinct phrases of $x^n$ parsed
jointly with the $k$--th distinct phrase $u(k)$ of $u^n$, $1\le k\le
c(u^n)$.\footnote{
Equivalently, $c_k(x^n|u^n)$ is the number of times $u(k)$ appears as a
parsed phrase of $u^n$.} The idea is that it is now the empirical {\it
conditional} distribution
of an $x$--phrase given a $u$--phrase that is uniform and is given
by $1/c_k(x^n|u^n)$ for all $c_k(x^n|u^n)$ $x$-phrases pertaining to $u(k)$.
For example,\footnote{The same example appears in \cite{Ziv85}.} if
\begin{eqnarray}
x^6&=&0~|~1~|~0~0~|~0~1|\nonumber\\
u^6&=&0~|~1~|~0~1~|~0~1|\nonumber
\end{eqnarray}
then $c(x^6,u^6)=4$,
$c(u^6)=3$,
$u(1)=\mbox{\tt `0'}$, $u(2)=\mbox{\tt `1'}$, $u(3)=\mbox{\tt `01'}$,
$c_1(x^6|u^6)=c_2(x^6|u^6)=1$, and $c_3(x^6|u^6)=2$.

Let us now
define, similarly as before:
\begin{equation}
\rho_E^\lambda(x^n|u^n)=\frac{c}{n\lambda}\log_2\left[\frac{1}{c}\sum_{i=1}^c
\exp_2\{\lambda L(y_{n_{i-1}+1}^{n_i})\}\right],~
~~~n_0\equiv 0,~~n_c\equiv n.
\end{equation}
As for a lower bound,
\begin{eqnarray}
\sum_{i=1}^c \exp_2\{\lambda L(y_{n_{i-1}+1}^{n_i})\}&=&
\sum_{k=1}^{c(u^n)}\sum_{i=1}^{c_k(x^n|u^n)} \exp_2\{\lambda
L(y_{n_{i-1}+1}^{n_i})\}\nonumber\\
&\ge&\frac{s^2}{2^{\lambda+1}-1}\cdot\sum_{k=1}^{c(u^n)}
\left(\exp_2\left\{(\lambda+1)\log\left[\frac{c_k(x^n|u^n)+s^2}{2s^2}\right]
\right\}-1\right),
\end{eqnarray}
which is, to the leading term (with respect to $\rho_E^\lambda(x^n|u^n)$), equivalent to 
\begin{equation}
\sum_{k=1}^{c(u^n)}\exp_2\{(\lambda+1)\log c_k(x^n|u^n)\}.
\end{equation}
An analogue of the alternative lower bound (\ref{alternativelb}) can also be
derived in the same way:
\begin{eqnarray}
\sum_{i=1}^c \exp_2\{\lambda L(y_{n_{i-1}+1}^{n_i})\}&=&
\sum_{k=1}^{c(u^n)}
\sum_{i=1}^{c_k(x^n|u^n)}
\exp_2\{\lambda L(y_{n_{i-1}+1}^{n_i})\}\nonumber\\
&\ge&\sum_{k=1}^{c(u^n)}
2^{(\lambda+1)\log c_k(x^n|u^n)-\lambda\gamma(s,\log_\alpha
c_k(x^n|u^n))]},
\end{eqnarray}
which is again, of the same asymptotic order.

For the upper bound, consider the joint incremental parsing of
$(x^n,u^n)$. For every $u(k)$, $k=1,2,\ldots,c(u^n)$, apply the LZ
algorithm separately, so that as before, the inner sum would contribute
\begin{equation}
\sum_{i=1}^{c_k(x^n|u^n)}
\exp_2\{\lambda L(y_{n_{i-1}+1}^{n_i})\}\le
(2\alpha)^{\lambda}2^{(\lambda+1)\log c_k(x^n|u^n)}.
\end{equation}
and so, overall, we get an upper bound of
\begin{equation}
(2\alpha)^{\lambda}\sum_{k=1}^{c(u^n)}\exp_2\{(\lambda+1)\log
c_k(x^n|u^n)\},
\end{equation}
which is asymptotically equivalent to both lower bounds in terms of the
achievability of $\rho_E^\lambda(x^n|u^n)$.

In this context, there is an interesting difference, that we observe, between the case without
side information, that was handled in Subsection 2.2, and the case with side
information considered here. While in the absence if side information, the empirical CGF agreed
with the ordinary LZ compressibility, $\frac{c\log c}{n}$ (see eq.\
(\ref{clogc})), here there is a difference between the empirical CGF, which is
roughly
\begin{equation}
\frac{c}{n\lambda}\log_2\left[\frac{1}{c}\sum_{k=1}^{c(u^n)}\exp_2\{(\lambda+1)\log
c_k(x^n|u^n)\}\right],
\end{equation}
and the ordinary LZ compressibility in the presence of side information
(see \cite{Ziv85}), which is about
$\frac{1}{n}\sum_{k=1}^{c(u^n)}c_k(x^n|u^n)\log c_k(x^n|u^n)$, the
natural individual--sequence analogue of the conditional entropy. 
Of course, the
latter can easily be recovered from the 
former by taking the limit $\lambda\to 0$.

\section{Summary and Conclusion}

In this work, we have made an attempt to develop complexity measures for
individual sequences that are analogous to the R\'enyi entropy of the
probabilistic case in the same spirit that the finite--state complexity
is analogous to the entropy rate of a stationary ergodic process.
We have examined 
both F--V and V--V definitions of the code--length CGF and obtained
different measures. In the F--V case, the main term was the R\'enyi
entropy derived from empirical distribution of non--overlapping blocks of the
given sequence, and an
alternative measure was given by the empirical CGF of $\{c_t\log c_t\}$,
where $c_t$ was defined as the number of distinct phrases in the $t$--block.
In the V--V version of the empirical CGF, the result actually coincides with
the ordinary complexity measure, $\frac{c\log c}{n}$. These findings were
finally extended to the setting of coding with side infotrmation at both
encoder and decoder, but in this case, there is a difference between the
empirical CGF and the ordinary conditional FS complexity.

%\section*{Acknowledgements}

%\section*{Appendix}
%\renewcommand{\theequation}{A.\arabic{equation}}
%    \setcounter{equation}{0}

\end{document}